# Meta-Data Objects as the Basis for System Evolution

Florida Estrella[1], Zsolt Kovacs[2], Jean-Marie Le Goff[2], Richard McClatchey[1] and Norbert Toth[1]

[1]*Centre for Complex Cooperative Systems, Univ. West of England, Frenchay, Bristol BS16 1QY UK*
[2]*CERN, Geneva, Switzerland*

*Abstract*

One of the main factors driving object-oriented software development in the Web-age is the need for systems to evolve as user requirements change. A crucial factor in the creation of adaptable systems dealing with changing requirements is the suitability of the underlying technology in allowing the evolution of the system. A reflective system utilizes an open architecture where implicit system aspects are reified to become explicit first-class (meta-data) objects. These implicit system aspects are often fundamental structures which are inaccessible and immutable, and their reification as meta-data objects can serve as the basis for changes and extensions to the system, making it self-describing. To address the evolvability issue, this paper proposes a reflective architecture based on two orthogonal abstractions - model abstraction and in-formation abstraction. In this architecture the modeling abstractions allow for the separation of the description meta-data from the system aspects they represent so that they can be managed and versioned independently, asynchronously and explicitly. A practical example of this philosophy, the CRISTAL project, is used to demonstrate the use of meta-data objects to handle system evolution.

Keywords: meta-objects, system evolution, self-describing objects, databases

# 1. Reflection

The capability of a system to reflect upon itself and be able to inspect and change its own state and behaviour is called reflection. A reflective system utilizes an open architecture where implicit system aspects are reified to become explicit first-class meta-objects [1]. These implicit aspects are often fundamental structures which are inaccessible and immutable. Meta-objects are the self-representations of the system describing how its internal elements can be accessed and manipulated. These self-representations are causally connected to the internal structures they represent, i.e. changes to these self-representations immediately affect the underlying system.

The use of reflection in computing creates a mutable and extensible system [2]. In a mutable system, the behaviour of existing constructs can be modified. On the other hand, an extensible system allows new features to be added. The ability to dynamically augment, extend and re-define system specifications can result in a considerable improvement in flexibility. This leads to dynamically modifiable systems which can adapt and cope with evolving requirements.

A reflective open architecture likewise increases a system's potential for reuse [3]. The customization mechanisms inherent in this architecture permit the system to be modified and reused for different needs. Making the components of the system self-representing or self-describing allows dynamic system re-configuration. It is therefore essential for such a self-describing system to have the capability to store descriptions about its dynamic structure, and make these descriptions available to the rest of the infrastructure as a consequence of how the system is connected.

The separation of system descriptions from the system aspects they represent is essential in the specification of evolvable OO system. To address the evolvability issue, this paper proposes a reflective architecture based on two orthogonal abstractions - model abstraction and information abstraction. In this architecture the modeling abstractions allow for the separation of description meta-data from the system aspects they represent so that they can be managed and versioned independently, asynchronously and explicitly.

The Meta-Object Protocol (MOP), a concrete manifestation of the open implementation technique, opens up language abstractions and its implementations to the programmers [1][4]. Consequently, programmers are capable of adjusting the language semantics and implementation better to suit their needs. Such flexibility in altering language semantics, and possibly improving performance through alternative implementation strategies, results in considerable benefit and program clarity for programmers in being able to customize the language semantics, and encourages programmers to participate in the language design process. MOP-based open languages are also called reflective programming languages. In general, a reflective system is a system which incorporates structures representing aspects of itself [5]. Such capability can only be attained if the language provides mechanisms which explicitly represent implicit aspects of the language itself, i.e. its descriptions and behaviour.

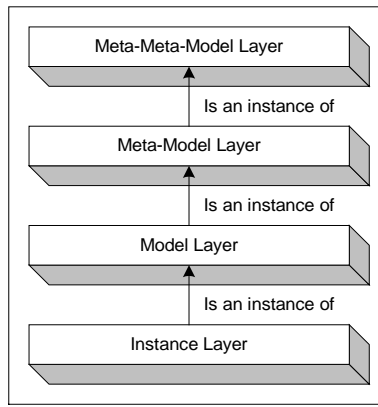

**Figure 1:** Four Layer Modeling Architecture

Some descriptions which play major roles in defining a language behaviour are class, attribute, association, inheritance, operation invocation, instance representation and the schema. To be able to dynamically modify these descriptions, they need to be turned into objects, thus creating class objects, attribute objects, association objects, etc. These objects are called meta-objects, as compared to application objects which are called base objects. Meta-objects control and manage the operations of base objects. The interface to these meta-objects is the MOP. In other words, the MOP is a set of operations used with meta-objects to define and configure system behaviour.

The promotion of implicit system descriptions to become explicit objects is called reification. The advantage of reifying system descriptions as meta-objects is that operations can be carried on them, like composing and editing, storing and retrieving, organizing and reading. Since these meta-objects represent system descriptions, their manipulation can result in change in system behavior. For reifying language descriptions like class, attribute and association, which themselves act as classes, what is needed is a mechanism for defining the class of a class. In OO programming, the class of a class object is a meta-class. Meta-objects, therefore, are implemented as meta-classes. The concept of meta-classes is a key design technique in improving the reusability and extensibility of these languages. VODAK [6], Prometheus [7], ADAM [8] and OM [9] are some of the next generation database management systems which have adopted the meta-class approach for tailoring the data model to adapt to evolving specifications. A meta-class may, typically, define properties about object creation, encapsulation, inheritance rules, message passing and the like.

## 2. Modeling Architectures

In modeling complex information systems, it has been shown that at least four modeling layers are required (see Figure 1) [10]. Each layer provides a service to the layer above it and serves as a client to the layer below it. The meta-meta-model layer defines the language for specifying meta-models. Typically more compact than the meta-model it describes, a meta-meta-model defines a model at a higher level of abstraction than a meta-model. Elements of the meta-meta-model layer are called meta-meta-objects. Examples of meta-meta-objects include MetaClass, MetaAttribute and MetaAssociation. These meta-meta-objects are also meta-classes whose instances are constructs corresponding to meta-model constructs.

The meta-model layer defines the language for specifying models. A meta-model is an instance of a meta-meta-model. It is also more elaborate than the meta-meta-model that describes it. Elements of the meta-model layer are called meta-objects, examples of which include Class, Attribute and Association. The model layer defines the language for specifying information domains. In this case, a model is an instance of a meta-model. Elements like Student, Teacher and Course classes are domain-specific examples of elements of the model layer. The bottom layer contains user objects and user data. The instance layer describes a specific information domain. Domain examples of user objects include the instances of Student, Teacher and Course classes. The Object Management Group (OMG) [11] standards group has a similar architecture based on model abstraction, with the Meta-Object Facility (MOF) model and the Unified Modeling Language (UML) [12] model defining the language for the meta-meta-model and meta-model layers, respectively.

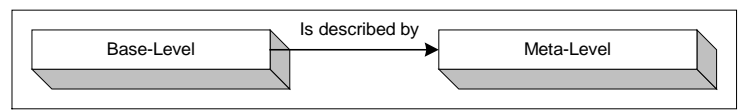

**Figure 2:** Meta-Level Architecture

Orthogonal to the model abstraction inherent in multi-layered meta-modeling approach is the information abstraction which separates descriptive information from the data they are describing. These system descriptions are called meta-data, as they are information defining other data. A reflective open architecture typifies this abstraction. A reflective open architecture is divided into two levels - the meta-level where the descriptive information reside and the base-level which stores the application data described by the meta-level elements. The meta-level contains the meta-data objects (also referred to as meta-objects in this paper) which hold the meta-data. These meta-objects manage the base-level objects. A two-layer architecture is shown in Figure 2.

The separation of meta-objects from base objects (see Figure 3) is essential in es-tablishing the difference between what an object does (in the base-level) from how it does it (in the meta-level). A meta-level architecture gives access to meta-objects and ensures that changes on the meta-objects lead to changes on the intended system as-pects represented by the meta-objects, i.e. the two levels are causally connected. For example, changes on the class Meta-object in the meta-level, via the class MOP, should result in the appropriate changes to all application objects in the base-level.

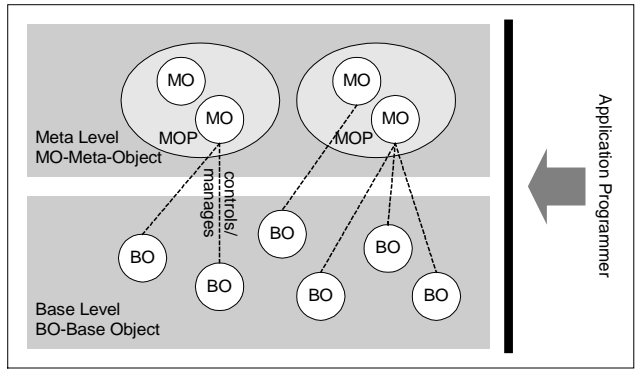

**Figure 3:** Separation of Meta-Objects from Base Objects

## 3. A Description-Driven Architecture

This paper proposes an architecture which combines the multi-layered meta-modeling approach with the meta-level architecture [13]. The description-driven architecture is illustrated in Figure 4. The layered architecture on the left hand side is typical of the layered systems and the multi-layered architecture specification of the OMG discussed earlier.

The relationship between the layers is Is an instance of. The instance layer contains data which are instances of the domain model in the model layer. Similarly, the model layer is an instance of the meta-model layer. On the right hand side of the diagram is another instance of model abstraction. It shows the increasing abstraction of information from meta-data to model meta-data, where the relationship between the two is Is an instance of as well. These two architectures provide layering and hierarchy based on abstraction of data and information models.

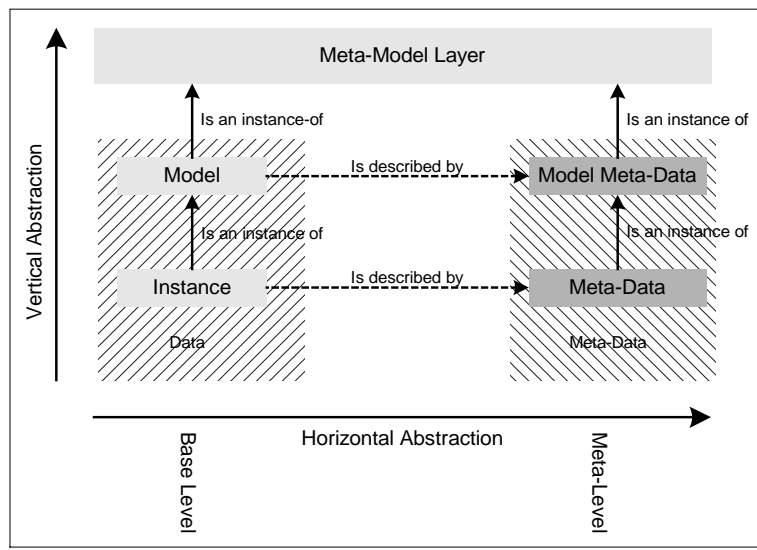

**Figure 4:** Description-Driven Architecture

The horizontal view provides an alternative abstraction where the relationship of meta-data and the data they describe are made explicit. This view is representative of the information abstraction and the meta-level architecture discussed earlier. The meta-level architecture is a mechanism for relating data to information describing data, where the link between the two is Is described by. As a consequence, the dynamic creation and specification of object types is promoted. The separation of system type descriptions from their instantiations allows the asynchronous specification and evolution of system objects from system types, consequently, descriptions and their instances are managed independently and explicitly. The dynamic configuration (and re-configuration) of data and meta-data is useful for systems whose data requirements are unknown at development time.

# 4. A Practical Example

This research has been carried out at the European Centre for Nuclear Research (CERN) [14] based in Geneva, Switzerland. CERN is a scientific research laboratory studying the fundamental laws of matter, exploring what matter is made of, and what forces hold it together. Scientists at CERN build and operate complex accelerators and detectors. Accelerators are huge machines to speed up particles very close to the speed of light, and then to let them collide with other particles. Detectors, on the other hand, are large instruments to observe what happens during these collisions.

The Compact Muon Solenoid (CMS) is a general purpose experiment that will be constructed from an order of a million parts and will be produced and assembled in the next decade by specialized centres distributed worldwide (see Figure 5). As such, the construction process is very data-intensive, highly distributed and ultimately requires a computer-based system to manage the production and assembly of detector components.

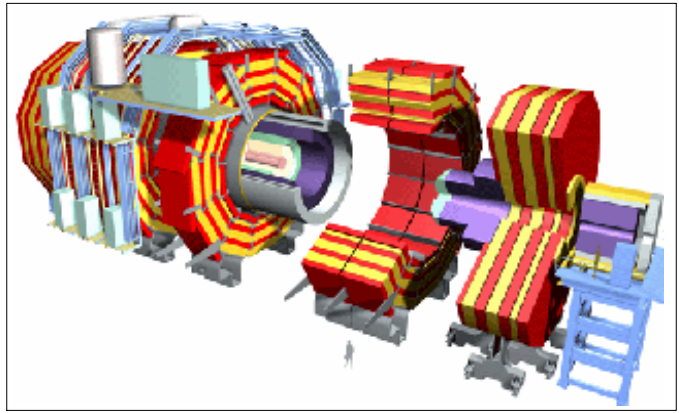

**Figure 5:** The CMS Detector

In constructing detectors like CMS, scientists require data management systems that are able of cope with complexity, with system evolution over time (primarily as a consequence of changing user requirements) and with system scalability, distribution and inter-operation. No commercial products provide the workflow and product data management capabilities required by CMS [15]. The design constraints imposed by CMS which are not currently satisfied by any commercial offering include:

- The workflow and product-related descriptions tend to evolve rapidly over time. The software must cater for the development of a high physics detector over an extended period of time (1999-2005) and whose design will naturally advance as time elapses. Hence the need to support long-running and potentially nested workflow activities, with natural consequences on transaction handling.
- The construction of CMS is one-of-a-kind. The evolution of workflows and product data must be allowed to take place as production continues. Consequently, versions of workflow and product descriptions coexist in the production process for the duration of the CMS construction. This is in contrast to industrial production lines where the process is seldom one-of-a-kind.
- The CMS construction is highly distributed. Production of (versions) of CMS products will take place in disparate areas all over the world. Each of these production centres must cater for multiple versions of evolving workflow and product descriptions in an autonomous manner but centrally coordinated from CERN.
- The data store must be reliably secure and available for a variety of purposes. Many users require different access to the CMS data, e.g. construction engineers interpret data using an assembly-oriented view, physicists view the detector data in terms of a set of electronically-decoded channels and mechanical engineers view the detector data in terms of constituent three-dimensional volumes aligned in space.

A research project, entitled CRISTAL (Cooperating Repositories and an Information System for Tracking Assembly Lifecycles) [16][17] has therefore been initiated, using OO computing technologies where possible, to facilitate the management of the engineering data collected at each stage of production of CMS. CRISTAL captures all the physical characteristics of detector components, which are, later, required by the physicists for activities such as detector construction, calibration and maintenance. CRISTAL is a distributed product data and workflow management system which makes use of an OO database for its repository, a multi-layered architecture for its component abstraction and dynamic object modeling for the design of the objects and components of the system. CRISTAL is based on a DDS using meta-objects. These techniques are critical to handle the complexity of such a data-intensive system and to provide the flexibility to adapt to the changing production scenarios typical of any research production system.

The design of the CRISTAL prototype was dictated by the requirements for adaptability over extended timescales, for system evolution, for inter-operability, for complexity handling and for reusability. In adopting a description-driven design approach to address these requirements, the separation of object instances from object descriptions instances was needed. This abstraction resulted in the delivery of a three layer description-driven architecture. The model abstraction (of instance layer, model layer, meta-model layer) has been adopted from the OMG specification, and the need to provide descriptive information, i.e. meta-data, has been identified to address the issues of adaptability, complexity handling and evolvability.

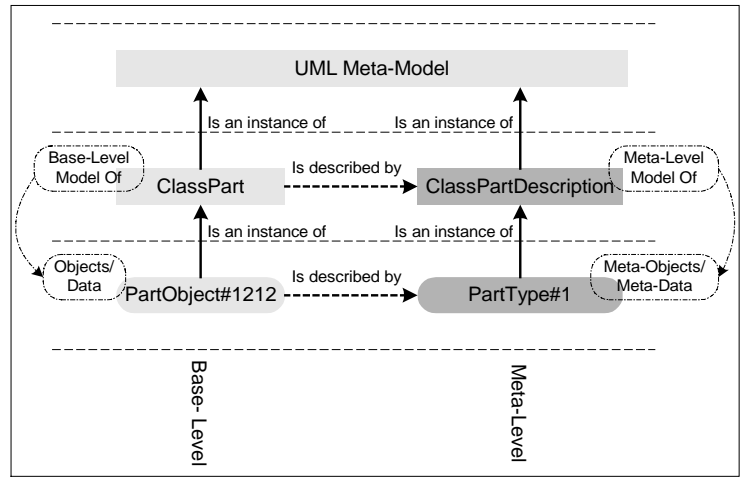

**Figure 6:** The CRISTAL Architecture

Figure 6 illustrates CRISTAL architecture. The CRISTAL model layer is comprised of class specifications for CRISTAL type descriptions (e.g. *PartDescription*) and class specifications for CRISTAL classes (e.g. *Part*). The instance layer is comprised of object instances of these classes (e.g. *PartType#*1 for *PartDescription* and *Part#1212* for *Part*). The model and instance layer abstraction is based on model abstraction and *Is an instance of* relationship. The abstraction based on meta-data abstraction and *Is described by* relationship leads to two levels - the meta-level and the base-level. The meta-level is comprised of meta-objects and the meta-level model which defines them (e.g. *PartDescription* is the meta-level model of *PartType#1* meta-object). The base-level is comprised of base objects and the base-level model which defines them (e.g. *Part* is the base-level model of the *Part#1212* object).

In the CMS experiment, production models change over time. Detector parts of different model versions must be handled over time and coexist with other parts of different model versions. Separating details of model types from the details of single parts allows the model type versions to be specified and managed independently, asynchronously and explicity from single parts. Moreover, in capturing descriptions separate from their instantiations, system evolution can be catered for while production is underway and therefore provide continuity in the production process and for design changes to be reflected quickly into production. As the CMS construction is one-of-a-kind, the evolution of descriptions must be catered for.

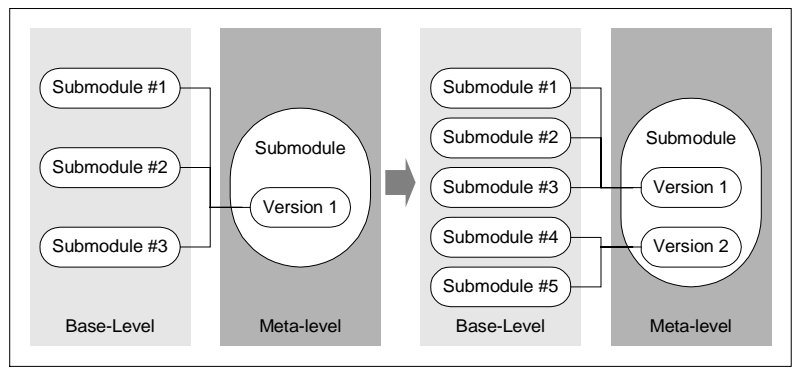

**Figure 7:** Evolving CMS Descriptions

The evolving CMS descriptions is illustrated in Figure 7. *Submodule Version 1* is a type object describing physical detector parts *Submodule#1*, *Submodule#2* and *Submodule#3*. *Submodule Version 2* (another type object) is a new version of the same type specification, and coexists with *Submodule Version 1* and its instances. In this example, *Submodule Version 1* and *Submodule Version 2* are instances of *ClassPartDescription* (in the meta-level). As new instances can be dynamically added into the system, consequently new versions and new type objects are handled transparently and automatically. In the base-level, *Submodule#1* is an instance of *ClassPart*. The *Is described by* relationship between the meta-level and the base-level elements allows for instantiations of physical parts to be described by versions of type objects in the meta-level. Hence, the separation of type descriptions in the meta-level from the data objects they describe caters for evolving system specifications.

## 5. Conclusions

The ubiquity of change in current information systems have contributed to the renewed interest in improving underlying system design and architecture. Reflection, meta-architectures and layered systems are the main concepts this paper has explored in providing a description-driven architecture which can cope with the growing needs of many computing environments. The description-driven architecture has two orthogonal abstractions combining multi-layered meta-modeling with open architectural approach allowing for the separation of description meta-data from the system aspects they represent. The description-driven philosophy facilitated the design and implementation of the CRISTAL project which required mechanisms for handling and managing evolving system requirements. In conclusion, it is interesting to note that the OMG has recently announced the so-called Model Driven Architecture as the basis of future systems integration [18]. Such a philosophy is directly equivalent to that expounded in this and earlier papers on the CRISTAL description-driven architecture.

### Acknowledgments

The authors take this opportunity to acknowledge the support of their home institutes. The support of P. Lecoq, J-L. Faure and M. Pimia is greatly appreciated. N. Baker, A. Bazan, T. Le Flour, S. Lieunard, L. Varga, G. Organtini and G. Chevenier are thanked for their assistance in developing the CRISTAL prototype.